\documentclass[prl,floatfix,showpacs,amsmath,twocolumn]{revtex4}
\usepackage{amssymb}
\usepackage{graphicx}
\usepackage{psfrag}

\begin{document}

\newcommand{\be}{\begin{eqnarray}}
\newcommand{\ee}{\end{eqnarray}}
\newcommand{\bea}{\begin{eqnarray}}
\newcommand{\eea}{\end{eqnarray}}
\newcommand{\bma}{\begin{subequations}}
\newcommand{\ema}{\end{subequations}}
\newtheorem{lemma}{Protocol}
\def\ket #1{\vert #1\rangle}
\def\bra #1{\langle #1\vert}

\def\lR{l^2_{\mathbb{R}}}
\def\RR{\mathbb{R}}
\def\E{\mathbf e}
\def\D{\boldsymbol \delta}
\def\S{{\cal S}}
\def\T{{\cal T}}
\def\dd{\delta}
\def\one{{\bf 1}}
\def\Flip{{\cal F}}
\def\1{{1}}
\def \eps {\varepsilon}

\title{Quantum simulations based on measurements and feedback control }

\author{K. G. H. Vollbrecht$^1$ and J. I. Cirac$^1$ }
\affiliation{$1$ Max-Planck Institut f\"ur Quantenoptik,
Hans-Kopfermann-Str. 1, Garching, D-85748, Germany }

\date{\today}

\begin{abstract}
We propose a scheme for performing quantum simulations with atoms in cavities based on a photon detection
feedback loop that requires only linear optical elements. Atoms can be stored individually without the need
of directly interacting  with one another. The scheme is able to
simulate any time evolution that can be written as a sum of two-qubit Hamiltonians, .e.g., any next neighbor interaction on a lattice.
It can also be made robust against photon losses.
\end{abstract}

\maketitle

The concept of building a quantum simulator goes back to the early days of
quantum information theory, where Feynman \cite{1} formulated the idea to
build one special quantum device to simulate another one.
This idea was forced by the impossibility to simulate any growing quantum system
efficiently on a classical device because of the exponentially growth of the number of
involved parameters. However, this restriction can be ruled out by any quantum device
due to the same exponentially behavior.
The remaining problem  is to  find a well controlled quantum system which can be forced to
evolve reliably under different time evolutions, especially under the
physically mostly relevant next neighbor Hamiltonians.

Several systems have been theoretically proposed  as candidates for quantum simulation and enormous
experimental effort has been put in the realization of the latter.
Atoms are one of the most promising candidates for storing a qubit, the basic quantum information
unit for the quantum simulator. Trapped in optical lattices \cite{OL} or arranged in
ions traps \cite{it1,it2} they allow for controllable next neighbor interactions that
enable the simulation of e.g., the Hubbard-model \cite{ja} or  spin-lattices \cite{ja2}.

Quantum simulations are closely related to the generation of entanglement since it is
the presence of entanglement that prevents the system to be simulated by classical algorithms.
Generation of entanglement between e.g. two atoms is usually achieved by either using direct interactions between
the atoms, or the interaction can be achieved indirectly by the exchange of photons \cite{ec}.

An alternative way to entangle two systems without ever getting into contact can be achieved by entanglement swapping.
Two photons emitted by the two atoms can be jointly measured \cite{ent2}. The outcome of the measurement defines a local feedback action
one has to apply to finish the entangling protocol.
This ansatz is avoiding  the difficulties that arise when a single photon has to be re-absorbed
by a single atom. Even in the case where we restrict to linear optical elements, entanglement can be reliable and efficient generated
in a probabilistic way using measurement and local feedback action \cite{ent2}.

Taking advantage of this idea for quantum simulation is hindered by two problems:
(i) Using measurements to run a quantum simulation is an unusual concept since this does not lead to an unitary time evolution.
The emitted photons have to be entangled with the atoms such that every measurement collapses the atom
state to a random outcome. Thus,  every single atom runs through its own random independent evolution and
it may be hard to synchronize all the atoms to simulate one controlled unitary evolution.
(ii) In a linear optical setup a complete Bell-measurement, as required for entanglement swapping, is not available \cite{ll}.
In consequence, the desired interaction can only be generated in a non-deterministic way.
 In addition,  a realistic photon is highly vulnerable to absorption and measurement losses.

In this letter we propose a scheme to circumvent those problems and to simulate effectively  arbitrary next neighbor interactions
among individually distance atoms with the help of emitted photons and measurements.	Although each
individual evolution is random, the system can be forced efficiently to follow any specific unitary next-neighbor
evolution.
The scheme can also be made resistant against any photon loss rates by keeping a backup copy of each photon.

\begin{figure}
    \psfrag{g0}[][]{\tiny $\ket{0}$}
	\psfrag{g1}[][]{\tiny$\ket{1}$}
    \psfrag{eo}[][]{\tiny$\ket{e_0}$}
	\psfrag{e1}[][]{\tiny$\ket{e_1}$}
    \psfrag{eo0}[][]{\tiny$\ket{e'_0}$}
	\psfrag{e11}[][]{\tiny$\ket{e'_1}$}

    \begin{center}
    	\includegraphics[height=2.5cm]{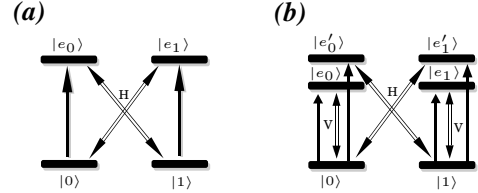}
    	\caption{The atom stores a qubit in two ground states. In addition several
    excited levels are used to create photons in the cavity. The double arrays indicate photon modes supported by the cavity, whereas
    the single arrows stand for external laser pulses to excite the atoms. In (a) only one photon mode is supported by the cavity. The photon
    qubit is encode is the presence or absence of a photon. In (b) two different photon modes are supported. }
    	\label{fig:lev1}
    \end{center}
\end{figure}

{\it  Photon generation:} We consider  atoms each in an optical cavity with the two ground states $\ket{0}$ and $\ket{1}$ realizing a qubit \cite{re}.
We assume that we can reliably apply any local unitary on the level structure of a single atom via lasers and external electromagnetic fields.
To couple two  atoms we need the atoms  to emit photons. For this purpose we need to include
more internal levels of the atoms.
We assume here two different possibilities depending on whether
we want to encode the photon-qubit into two different polarizations or the photon occupation number (photon/ no photon) :

{(a) \it Photon/no photon:}
The atom has the two excited levels $\ket{e_0}$ and $\ket{e_1}$, which decay via the same cavity mode
into  the ground state as shown  in Figure \ref{fig:lev1} (a).
The two ground states can simultaneously be changed in the two excited state by a laser pulse. By changing the strength
of the laser pulse we can bring the ground states $\ket{0}, \ket{1}$
into any  superposition of the ground and the excited states, i.e.,
$\ket{0}\rightarrow \sqrt{1-\eps} \ket{0}+\sqrt{\eps} \ket{e_0}$ and  $\ket{1}\rightarrow \sqrt{1-\eps} \ket{1}+\sqrt{\eps} \ket{e_1}$.
When the excited state decays the atom emits a photon such that we get the transformation
\bea \label{eq:1}
\ket{0}&&\rightarrow \sqrt{1-\eps} \ket{0}\ket{V}+\sqrt{\eps} \ket{1}\ket{H}, \\ \nonumber
\ket{1}&&\rightarrow \sqrt{1-\eps} \ket{1}\ket{V}+\sqrt{\eps}\ket{0}\ket{H}
\eea
where $\ket{H}$ denotes a photon in the cavity mode, while $\ket{V}$
stands for the vacuum, i.e.,  the absence of any photon.

{(II) \it Polarized photons:}
A similar kind of transformation can be archived using the level structure showed
in Figure \ref{fig:lev1}(b).
The atom has four excited levels and the cavity supports two different cavity modes, called
$H$ and $V$. If the atom is in level $\ket{e_0} (\ket{e_1}$) it decays into $\ket{0}$ ($\ket{1}$) and
emits a $V$-photon to the cavity, while when in state $\ket{e'_0}$ ($\ket{e'_1}$) it decays to
$\ket{1}$ ($\ket{0}$) and emits a $H$-photon.
With two laser pulses we can excite the ground states into
 a superposition of the four excited states
$\ket{0}\rightarrow \sqrt{1-\eps}\ket{e_0}+\sqrt{\eps} \ket{e'_0}$ and  $\ket{1}\rightarrow \sqrt{1-\eps}\ket{e_1}+\sqrt{\eps} \ket{e'_1}$,
where the $\eps$ can be controlled by the strength of the  lasers.
After the  atom has decayed back into the ground states  we get a transformation similar to (\ref{eq:1}) with
the only difference that $\ket{V}$ stands now for a photon and not for the vacuum.

In both cases the original qubit encoded in the ground state  gets flipped  with probability $\eps$ where the flip can be
recognized by a measurement of the emitted photon.
This basic operation will be denoted as $U_\eps$.

\begin{figure}
    \begin{center}
    	\includegraphics[height=2.5cm]{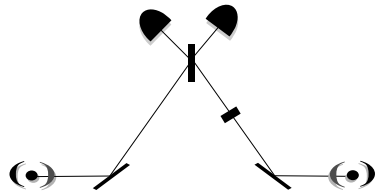}
    	\caption{Two distance atoms are effectively couplet by sending photons coming out of the cavity to a beam splitter.
         Dependent on the measurement outcome we apply a feedback loop on the atoms until the desired
         evolution has been done.}
    	\label{fig:2}
    \end{center}
\end{figure}

{\it Feedback loop:}
To get an effective interaction between  two atoms we chose the following strategy:
we apply $U_\eps$ to the first atom and $U_{(1-\eps)}$ followed by a flip $F$ to the second atom.
The flip operations exchanges the states $\ket{0}$ and $\ket{1}$.
The two (possible) photons emitted by the two atoms are sent onto a beam-splitter and measured.
At one branch we add an extra phase of $i$ to the $H$-photon (see Fig. \ref{fig:2}).

If $\ket{\psi}$ denotes the  initial state of the two atoms
we will get after emitting the photons the state
\bea\label{eq:state}
U_\eps \otimes F U_{(1-\eps)} \ket{\psi} =\\\nonumber
(1-\eps) \ket{\psi} \ket{V H}\\\nonumber
+i \eps \sigma \otimes \sigma \ket{\psi} \ket{H V}\\\nonumber
+\sqrt{\eps(1-\eps)} 1\otimes \sigma \ket{\psi} \ket{V V}\\ \nonumber
+i \sqrt{\eps(1-\eps)} \sigma \otimes 1 \ket{\psi}\ket{H H},
\eea
where $\sigma$ is the first pauli matrix that flips the two states $\ket{0}, \ket{1}$.
After passing the beam-splitter  the photons are measured in the $H,V$ basis. This correspond to an incomplete Bell-measurement of
the two-photon pair, i.e., they are projected in one of the four states
$\ket{-}:=\ket{H V- V H},\ket{+}:=\ket{H V +  V H}, \ket{ H H},\ket{ V V}$. The
remaining two atoms are projected onto
\begin{itemize}
\item[(i)]
 $\ket{-}\mapsto (1-\eps) \ket{\psi}-i \eps \sigma \otimes \sigma \ket{\psi}$ with probability $\frac{1}{2}((1-\eps)^2+\eps^2)$;
\item[(ii)] $\ket{+}\mapsto (1-\eps) \ket{\psi}+i \eps \sigma_ \otimes \sigma\ket{\psi}$ with probability $\frac{1}{2}((1-\eps)^2+\eps^2)$;
\item[(iii)] $\ket{VV}\mapsto(1 \otimes \sigma)\ket{\psi}$ with probability $\eps(1-\eps)$;
\item[(iv)] $\ket{HH}\mapsto(\sigma \otimes 1)\ket{\psi}$ with probability $\eps(1-\eps)$.
\end{itemize}

Let us first assume that the only possible  outcomes would be $\ket{+},\ket{-}$.
In both cases the resulting state can be written as $e^{i t \sigma \otimes \sigma} \ket{\psi}=\cos{(t)} \ket{\psi} + i \sin{(t)}\sigma \otimes \sigma \ket{\psi} $ with $t$ equal
to  $t=\pm \arcsin(\eps)\approx \eps$. So we can simulate any time evolution with respect to
the Hamiltonian $\sigma \otimes \sigma$ over a time $t=\pm \eps$, but with a random time direction, i.e.,
by repeating we can simulate a random walk on the time axis of the evolution. If we aim for
a time simulation $V(t)=e^{i t \sigma \otimes \sigma}$ we can repeat
the protocol until the random walk by chance matches the desired time $t$. We can speedup this strategy
 by a feedback loop controlling the $\eps$ parameter.
In a first step we choose $\eps$ in such a way, that $\sin{t}=\eps$ giving us a chance of $1/2$
to succeed with $V(t)$ in the first step.
We will  assume $t$ to be small such that we can  approximate $\eps=t$ in the following discussion, but in principle everything can
be done exactly. In a case of a failure $\ket{-}$ we  end up with $V(-t)$.  By increasing $\eps$ to $2 \eps$ we apply randomly $V(\pm 2 t)$ in the next step such that in the successful
$\ket{+}$-case we compensate the previous $V(-t)$ rotation and in addition are left with the desired
$V(t)$ rotation. In case of further failures  we each time double $\eps$  and retry. This way we guarantee for a constant success probability
of $1/2$ in each round and
we can simulate every evolution
$V(t)$
with exponential increasing probability in the number of rounds.


Problems may  occur if we measure $\ket{H H}, \ket{V V}$ in between. But in these cases
we applied either $1 \otimes \sigma$ or $\sigma \otimes 1$ which can be corrected by local gates.
 Since our Hamiltonian $\sigma \otimes \sigma$
commutes with these unitaries we can ignore these measurement results and just correct the overall error in the end. Note that
this overall error only can be one of the three local unitaries  $1 \otimes \sigma, \sigma \otimes 1, \sigma \otimes \sigma$.
Since the probability of $\ket{+},\ket{-}$ is at least $1/4$ in each round the success probability is still exponentially good.
So we can efficiently simulate $V(t)$ for arbitrarily $t$ up to some random but known Pauli-errors that commute with the evolution.

{\it Simulations for $H=\sigma_k \otimes \sigma_l $:} Given $V(t)$ and the possibility to do local unitary gates at each single atom, we can apply
\bea
V^{kl}(t)=e^{i t \sigma_k \otimes \sigma_l}\approx (\1 + i t \sigma_k \otimes \sigma_l)
\label{eq:3}
\eea
for arbitrarily Pauli matrices $\sigma_k \otimes \sigma_l$, since $V^{kl}(t)= u_k \otimes u_l V(t) u_k^\dagger \otimes u_l^\dagger$ with
the local unitaries defined by $u_k = \sigma_k^{\frac{1}{2}} \sigma_1^{-\frac{1}{2}}$. If the measurement outcome is $\ket{HH}, \ket{VV}$ we pick up an error
of $1 \otimes \sigma_l$ or $\sigma_k \otimes 1$ which commutes with (\ref{eq:3}).

{\it Simulation of an arbitrarily two-qubit Hamiltonian:}
Every two-qubit Hamiltonian can be written as $H=\sum_{kl} \lambda_{kl} \sigma_k \otimes\sigma_l$.
By successively applying $V^{kl}(\eps \lambda_{kl})$ we can simulate
$$\prod_{kl} V^{kl}(\eps  \lambda_{kl})= \1 + i \eps \sum_{kl} \lambda_{kl} \sigma_k \otimes \sigma_l+O(\eps^2)$$
which equals $e^{i \eps H}$ up to first order in $\eps$. Note that each
$ V^{kl}(\eps  \lambda_{kl})$ can produce an error that does not commute with the following
$ V^{k'l'}(\eps  \lambda_{k'l'})$. But since only products of Pauli-matrices are involved,
the possible errors either commute or anti-commute with the following evolutions.  In the case that
$V^{k'l'}$ anti-commute only  the time direction is inverted which we
easily can compensate by  exchanging the $\ket{+}$ and the $\ket{-}$ case in above protocol. So we
can simulate any product  $\prod_{kl} V^{kl}$ up to a random but known error $\sigma_r \otimes \sigma_p$, i.e.,
$\ket{\psi} \rightarrow \sigma_r \otimes \sigma_p  \prod_{kl} V^{kl} \ket{\psi}.$
\begin{figure}
    \begin{center}
    	\includegraphics[height=2.5cm]{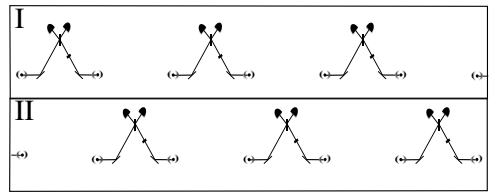}
    	\caption{To simulate a next neighbor Hamiltonian we can do $m/2$ of the rotations in parallel. We need
    to basic setup such that each cavity sends a photon on a beam splitter together with either its right or its left neighbor.  }
    	\label{fig:sim}
    \end{center}
\end{figure}
This allows us to take advantage of the
Trotter-formula \cite{sl} to approximate $e^{i  t H}$ by
$$\lim_{n \rightarrow \infty}\left[ \prod_{kl} V^{kl} \left( \frac{t}{n}  \lambda_{kl}\right) \right]^n= e^{i  t H}$$
with a precession of the order of $\frac{1}{n}$.

{\it Simulation of an arbitrarily next neighbor Hamiltonian:}
In the same manner we can simulate any next-neighbor Hamiltonian on a lattice, e.g. the 1D Hamiltonian
$H=\sum_{kl, x=1}^{x=m} \lambda_{kl}^{(x)} \sigma^{(x)}_k \otimes\sigma_l^{(x+1)}$, where $x$ labels the different qubits.
Using the Trotter-formula and the $V^{kl}$ we can approximate $e^{i  t H}$ by
$$\lim_{n \rightarrow \infty}\left[ \prod_{klx} V_x^{kl} \left( \frac{t}{n}  \lambda_{kl}\right) \right]^n= e^{i  t H}$$
with a precession of the order of $\frac{m}{n}$ up to an overall error that is just a product of random but known Pauli-matrices.
The same approximation can be made for any lattices with next-neighbor interactions or for any arbitrary Hamiltonian consisting
 of sums of two-qubit Hamiltonians.

{\it Time estimation: }
Assume we want to simulate a Hamiltonian $H=\sum_{i=1}^{m'} H_i$, where $H_i$ denotes a two qubit Hamiltonian.
Using the Trotter formula we can approximate the time evolution $e^{i H t}$ by applying $4 m n$  $V^{kl}$-rotations.
 To get a constant precision we have to scale $n$  at least linearly with $m$  such that
  we  need  $O(m^2)$ applications of a single $V^{kl}$-rotation for a proper simulation.
  For every $V^{kl}$-rotation we need one successful application of the feedback loop
which succeeds with a probability of about $\frac{1}{2}$ (or at least $\frac{1}{4}$ ). If we
apply $2 m^2+ \frac{const}{2} m$ steps we get at least  $m^2$ successfully feedback loops with a probability that exponentially
approaches $100 \%$, e.g. $97\%$ for $const=3$. Therefore we can simulate any Hamiltonian with $m$ two-qubit terms in
a time that scales  quadratically with $m$.

In a $1D$ $m$-particle  next neighbor setups the Hamiltonian will
consist out of $m$ terms. Since only two qubits
are involved in each local term we can speed up the simulation by  applying at a time $m/2$ of  the steps in parallel (see Fig. \ref{fig:sim}).
Assume we want to make $m/2$ gates in parallel in a time that allows for $c \log(m)$ feedback loops.
Then each single gate succeeds  with $(1-2^{-c \log(m)})=(1-m^{-c \log(2)})$.
So the total probability of success of the $m/2$ gates is $(1-m^{-c \log(2)})^{m/2}$ which approaches $1$ as long as
$c\log(2)>1$.   This whole procedure has to be repeated $2 m$ times to do the required $m^2$ $V^{kl}$-rotations.
Thus,  we can simulate with arbitrarily high probability $m$ particles in a 1D setup in a time scaling like  $O(m\log(m))$.

{\it Errors:}
The protocol proposed so far is highly vulnerable to photon losses and detection losses. A photon loss is especially fatal
if we miss to identify a $\ket{HH}$ or $\ket{VV}$ measurement result, because we miss to identify a Pauli-error.

In the case we work with polarized $H$ and $V$ photons we know whether we missed to detect a photon or not, since
there are always two photons generated.
We can use this fact to fight photon losses by the following idea. We will keep a backup-copy of the photon
that we can use in the case that the real photon is lost. To store this copy assume a second 'atom' is present
in each cavity.
So we have
two atoms (called A and B atom)  with two internal levels each and we assume that we can apply arbitrarily two qubit gates between them.
 In the first step the $B$-atoms  will play  now the role of the photons to get the backup-copy before the photon is even generated:
We start with the two atom state $\ket{\psi}_{AA'}$ and the two extra atoms in state $\ket{00}_{BB'}$ and apply $U_\eps$ now between
the two atoms A and B (and A' and B'), i.e.,
$\ket{0}_A\ket{0}_B \rightarrow \sqrt{1-\eps} \ket{0}_A\ket{0}_B+ \sqrt{\eps}  \ket{1}_A\ket{1}_B$ and
 $\ket{1}_A\ket{0}_B \rightarrow \sqrt{1-\eps} \ket{1}_A\ket{0}_B+ \sqrt{\eps}  \ket{0}_A\ket{1}_B$.
 After applying this we are left with a state of form (\ref{eq:state}) where the photon modes $V, H$ are replaced by
 the internal states $\ket{0},\ket{1}$ of the two $B$-atoms. The next step is to copy the internal states
 of the $B$-atoms onto photons, such that we exactly get (\ref{eq:state}) and can continue the protocol.

 To this end we have to  force the two $B$-atoms to produce a photon dependent on their internal level, i.e.,
$\ket{0}_B \rightarrow \ket{0}_B\ket{V}, \ket{1}_B \rightarrow \ket{1}_B\ket{H} $.
Note that the B-atom will keep
 their quantum information  and can be reused in the case of a photon loss.
  If we measure both $B$-atoms  in the $\ket{0+1}, \ket{0-1}$ basis then we  destroy this information and are left with a state similar to (\ref{eq:state}) up to some random but known sign-phases that will change only the time directions. So we can use the above protocols up to the fact, that we have to include the measurement results of the $B$-atoms to identify the direction of time evolution.

Instead of measure the 'B' atoms  in the $\ket{0+1}, \ket{0-1}$ basis we can measure them individually in the standard basis  $\ket{0}, \ket{1}$.
In this case $\ket{\psi}$ stays unchanged up to one of the random but known errors  $1 \otimes \sigma_l, \sigma_k \otimes 1, \sigma_k \otimes \sigma_l$. We do not get any time evolution, but are still able to continue the protocol.

The main idea is now to first measure the photons and then the atoms.
If both photons arrived we continue by measuring the $B$-atoms in the  $\ket{0+1}, \ket{0-1}$ basis.
If one or both photons are missing we measure in the   $\ket{0}, \ket{1}$ basis  to protect our system against random unknown errors.
Note that this measurement can be repeated until we successfully detected a photon.
Doing so, we can handle any photon loss rates.

\begin{figure}
    \psfrag{g0}[][]{\tiny $\ket{00}$}
	\psfrag{g1}[][]{\tiny$\ket{01}$}
    \psfrag{h0}[][]{\tiny $\ket{10}$}
	\psfrag{h1}[][]{\tiny$\ket{11}$}
    \psfrag{eo}[][]{\tiny$\ket{e_0}$}
	\psfrag{e1}[][]{\tiny$\ket{e_1}$}
    \psfrag{eo0}[][]{\tiny$\ket{e'_0}$}
	\psfrag{e11}[][]{\tiny$\ket{e'_1}$}

    \begin{center}
    	\includegraphics[height=2.5cm]{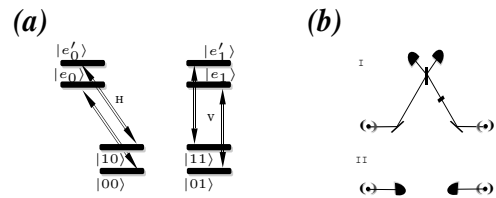}
    	\caption{(a) to fight photon losses the atom needs to store a second qubit to carry a security copy of the photon.
    (b) in the first attempt we try to establish an evolution by sending the photons on a beam-splitter. In case of success
    we erase the security copy, in case of failure we can reuse the copy by measuring in different basis. }
    	\label{fig:lev8}
    \end{center}
\end{figure}

Instead of putting two atoms in a cavity we can as well use one single atom with four internal levels. In the four internal levels
we can encode the $A$ and the $B$ atom.
To this end we need an internal level structure that allows for storing two-qubits and the possibility to measure one of this
qubits independently from the other. On possible level structure is shown in Fig. \ref{fig:lev8}. The cavity supports two
photon modes. If we excite $\ket{10},\ket{00}$ to $\ket{e'_0},\ket{e_0}$ and $\ket{11},\ket{01}$ to $\ket{e'_1},\ket{e_1}$, we can
measure the second qubit while the first qubit stays untouched.

{\it Quantum computation: }
Note that the  same kind of setup can be used to do a quantum computation. If we apply $V(t)$ to two atoms with $t$ equal to
$\pi/2$ we get a two-qubit gate that is locally equivalent to a c-not gate. Together with local gates on single atoms this allows for universal quantum computation. This scheme is similar to the so called 'repeat until success'-schemes proposal in \cite{Almut}.
The advantage we have here is that in our proposal a single two-qubit gate can be made photon loss resistant, whereas in
\cite{Almut} photon losses can only be attacked via the creating of cluster states which then are used as a resource for
one way quantum computation.

{\it Conclusion:}
We have proposed a scheme for performing quantum simulations  based on a photon detection
feedback loop that requires only linear optical elements. Atoms can be stored individually without the need
to directly realize an interaction between two atoms. The scheme can
simulate any time evolution that can be written as a sum of two-qubit Hamiltonian in a time scaling polynomially with the number
of particles. Our scheme can be modified to be resistant to any amount of photon losses by storing two qubit per cavity, whereas
the second qubit carries a backup-copy of the photons.
 The same kind of setup can be used
to do c-not gates and implement a universal quantum computer.
So far the simulation schemes needs several discrete round.
One challenge would be to make the interaction and the feedback control continuous \cite{me} and establish
a continuous feedback theory for a quantum simulator.
We
acknowledge support from   DFG-Forschungsgruppe 635 and SCALA.


\begin{thebibliography}{99}
\bibitem{1} R. P. Feynman, Opt. News 11, 11 (1985): Found. Phys.
16 507 (1986); Int. J. Theor. Phys. 21, 467 (1982).

\bibitem{OL} M. Greiner, O. Mandel, T. Esslinger, T.W. H\"ansch, I.
Bloch, Nature(London) 415, 39 (2002).


\bibitem{it1} D. Leibfried, R. Blatt, C. Monroe, D. Wineland,
Rev.Mod. Phys. 75, 281 (2003).
\bibitem{it2} D. Porras, J.I. Cirac, 	Phys. Rev. Lett. 92, 207901 (2004);
	Phys. Rev. Lett. 93, 263602 (2004); 	Phys. Rev. Lett. 96, 250501 (2006).

\bibitem{ja} D. Jaksch, C. Bruder, J. I. Cirac, C. W. Gardiner, and
P. Zoller, Phys. Rev. Lett. 81, 3108 (1998).
\bibitem{ja2} L.-M. Duan, E. Demler, and M. D. Lukin, Phys. Rev.
Lett. 91, 090402 (2003).

\bibitem{re} A. Kubanek, A. Ourjoumtsev, I. Schuster, M. Koch, P.W.H. Pinkse, K. Murr, G. Rempe, arXiv:0811.0264 ;
A. D. Boozer, A. Boca, R. Miller, T. E. Northup, H. J. Kimble, Phys. Rev. Lett. 98 (2007) 193601 .



\bibitem{ll} N. Luetkenhaus, J. Calsamiglia, and K.A. Suominen Phys. Rev. A 59 3295 (1999)

\bibitem{ec} S.J. van Enk, J.I. Cirac, P. Zoller, Phys.Rev.Lett. 78 (1997) 4293-4296



\bibitem{ent2}  C. Cabrillo, J. I. Cirac, P. Garc´ýa-Fern´andez, and
P. Zoller, Phys. Rev. A 59, 001025 (1999); . Bose, P. L. Knight, M. B. Plenio, and V. Vedral, Phys.
Rev. Lett. 83, 5158 (1999).


\bibitem{sl} Seth Lloyd, 	Science 273, 1073 (1996).
\bibitem{Almut} Yuan Liang Lim, Almut Beige, Leong Chuan Kwek, 	Phys. Rev. Lett. 95, 030505 (2005);
S. C. Benjamin, D. E. Browne, J. Fitzsimons, J. J. L. Morton,  New Journal of Physics 8, 141 (2006).
\bibitem{me} K.G. Vollbrecht et.al. , in preparation (2008).
\end{thebibliography}
\end{document}